\documentclass[11pt,twoside]{article}

\usepackage{asp2004}
\usepackage{epsf}
\usepackage{psfig}
\usepackage{lscape}

\def\hi{{\sc H~i}}

\begin{document}
\title{Multiscale Structure in Dust Reflection and Cold \hi}  
\author{Steven J. Gibson}   
\affil{National Astronomy and Ionosphere Center}    

\begin{abstract} 
I present 2-D angular power spectra of cold \hi\ emission and optical dust
reflection tracing the \hi\ in the Pleiades reflection nebula.  This analysis
reveals a uniform power-law slope of $-2.8$ over 5 orders of magnitude in
scale, from tens of parsecs down to tens of astronomical units.
\end{abstract}



\section{Introduction}

Multiscale structure in the neutral atomic ISM is well known (e.g.,
\citealt{g93,ddg00,d01}).  A quantitative examination of this structure over a
wide range of scales may help us understand its
physical causes (e.g.\ turbulent cascades), especially if the smaller scales
can be probed.  Such an examination is usually difficult with \hi\ 21cm line
observations, due to sightline confusion and sensitivity limits at high
resolution.  However, \hi\ and dust should be well mixed in cold gas, so dust
may 
trace \hi\ structure at fine scales.  Power-spectrum studies of dust infrared
emission show power-law behavior similar to \hi\ emission (e.g.,
\citealt{g92}), and one study of optical dust reflection in cirrus does as well
\citep{gc94}.

\begin{figure}[!ht]
\plottwo{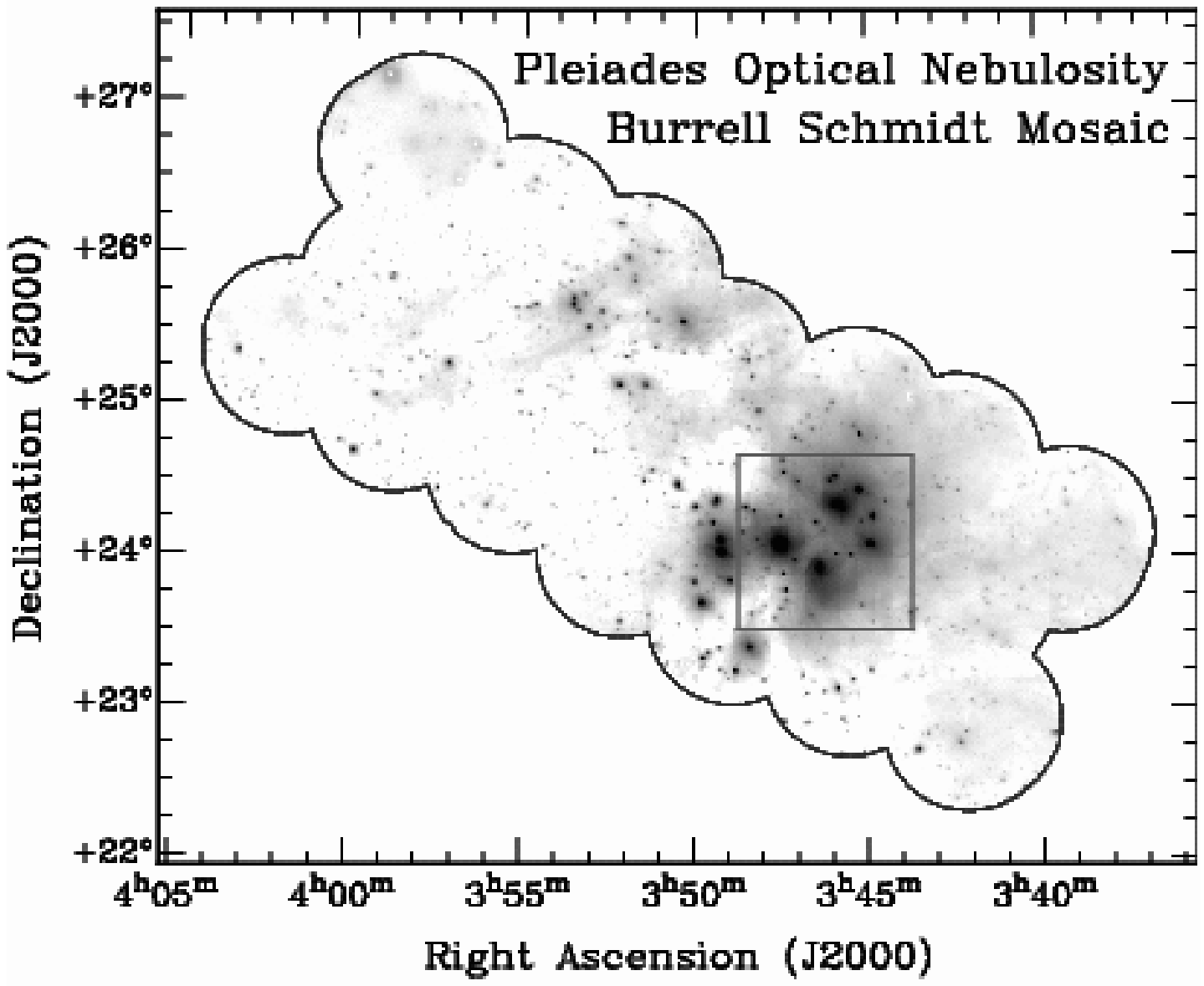}{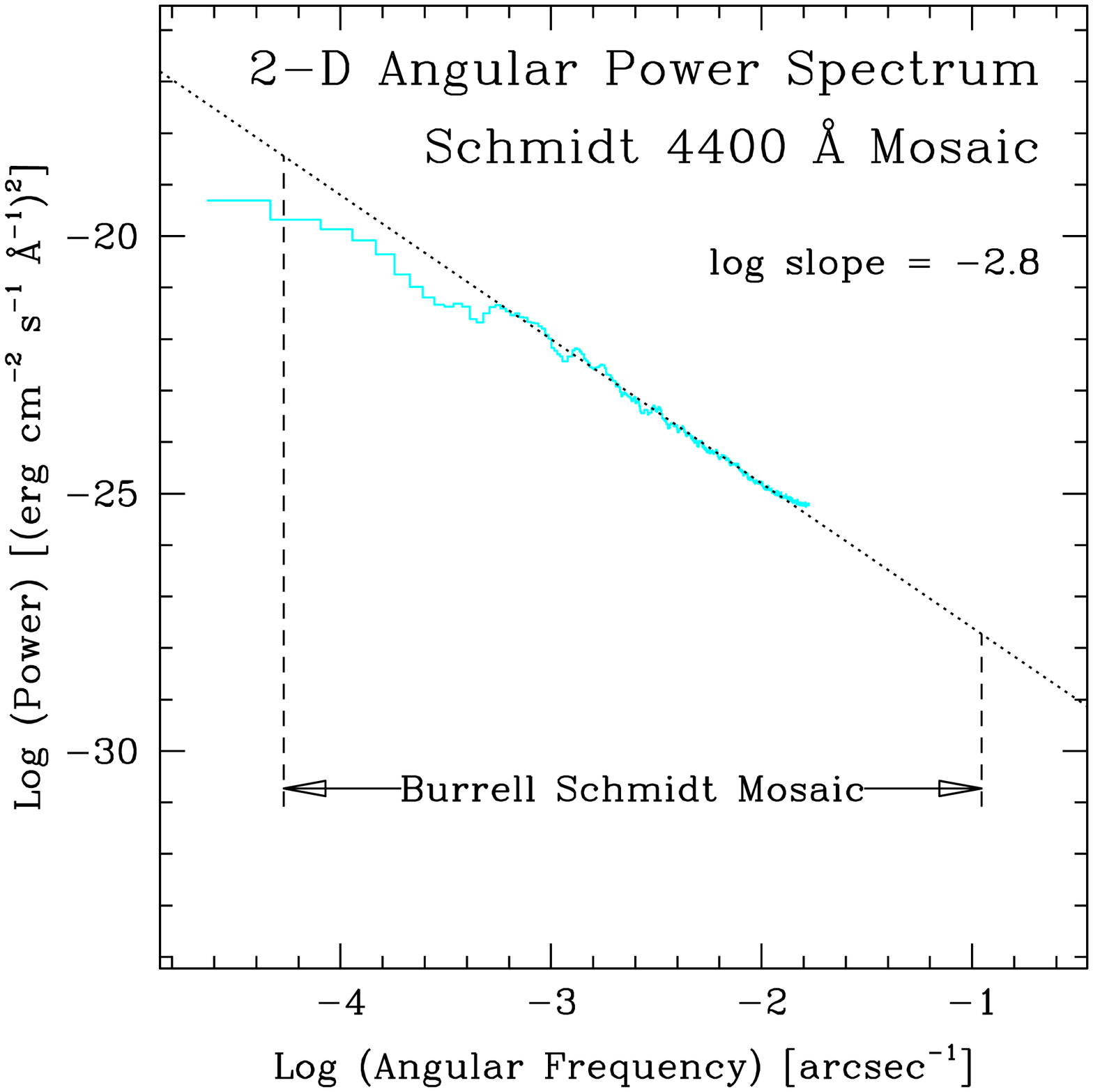}
\plottwo{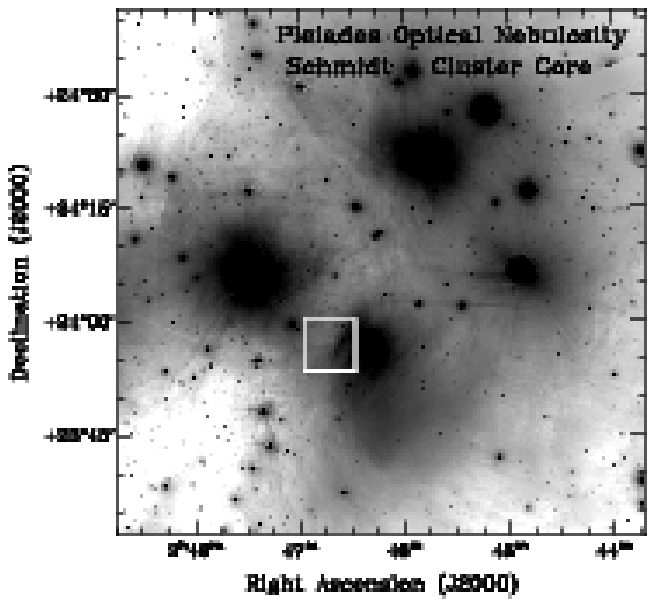}{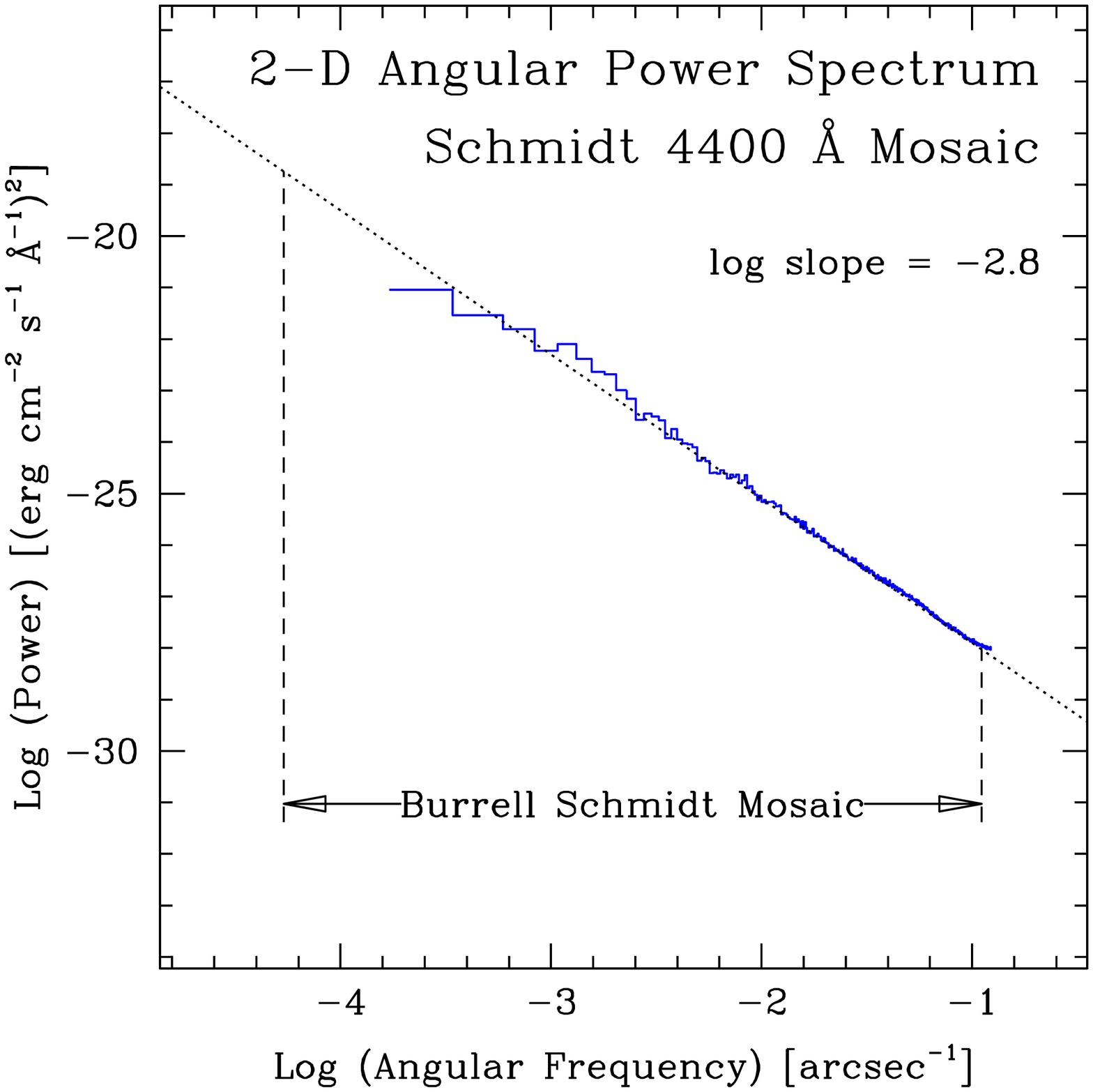}
\caption
{
Pleiades optical nebula, negative logarithmic intensity scale.  
{\itshape Top left:\/} 40-field mosaic from the 0.6m Burrell Schmidt telescope
\citep{gn03a}.  The boxed area is shown in the lower panel.
{\itshape Bottom left:\/} nebular core in the same mosaic.
The boxed area is shown in Figure \ref{Fig:wiyn}.
{\itshape Right panels:\/} 2-D angular power spectra derived from the images
at left.  
}
\label{Fig:schmidt}
\end{figure}

\begin{figure}[!ht]
\plottwo{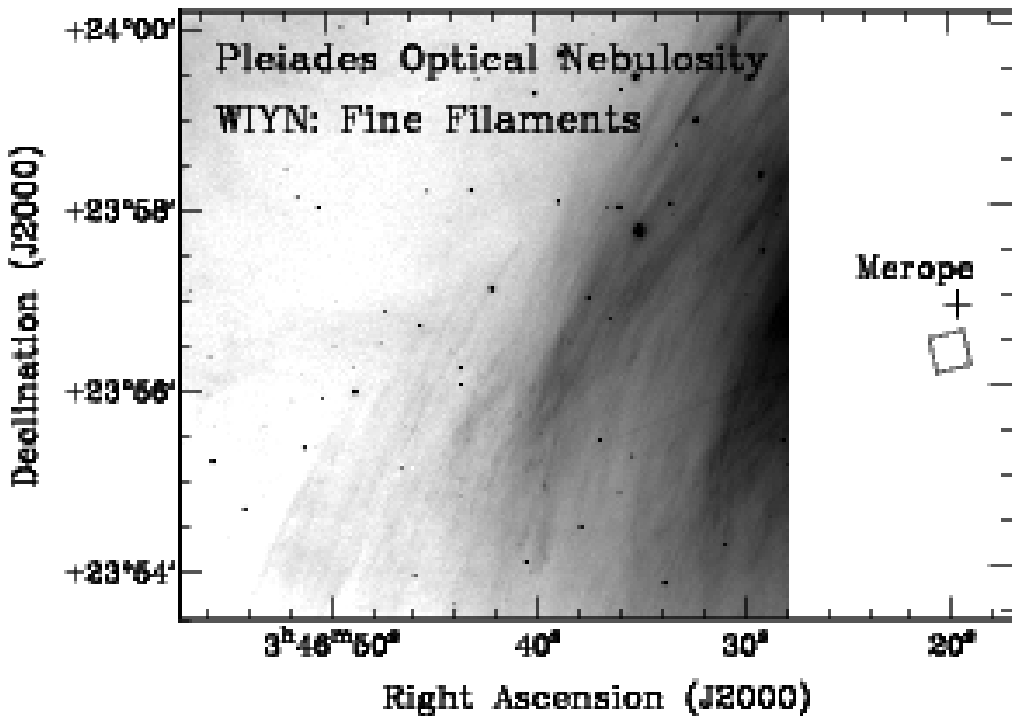}{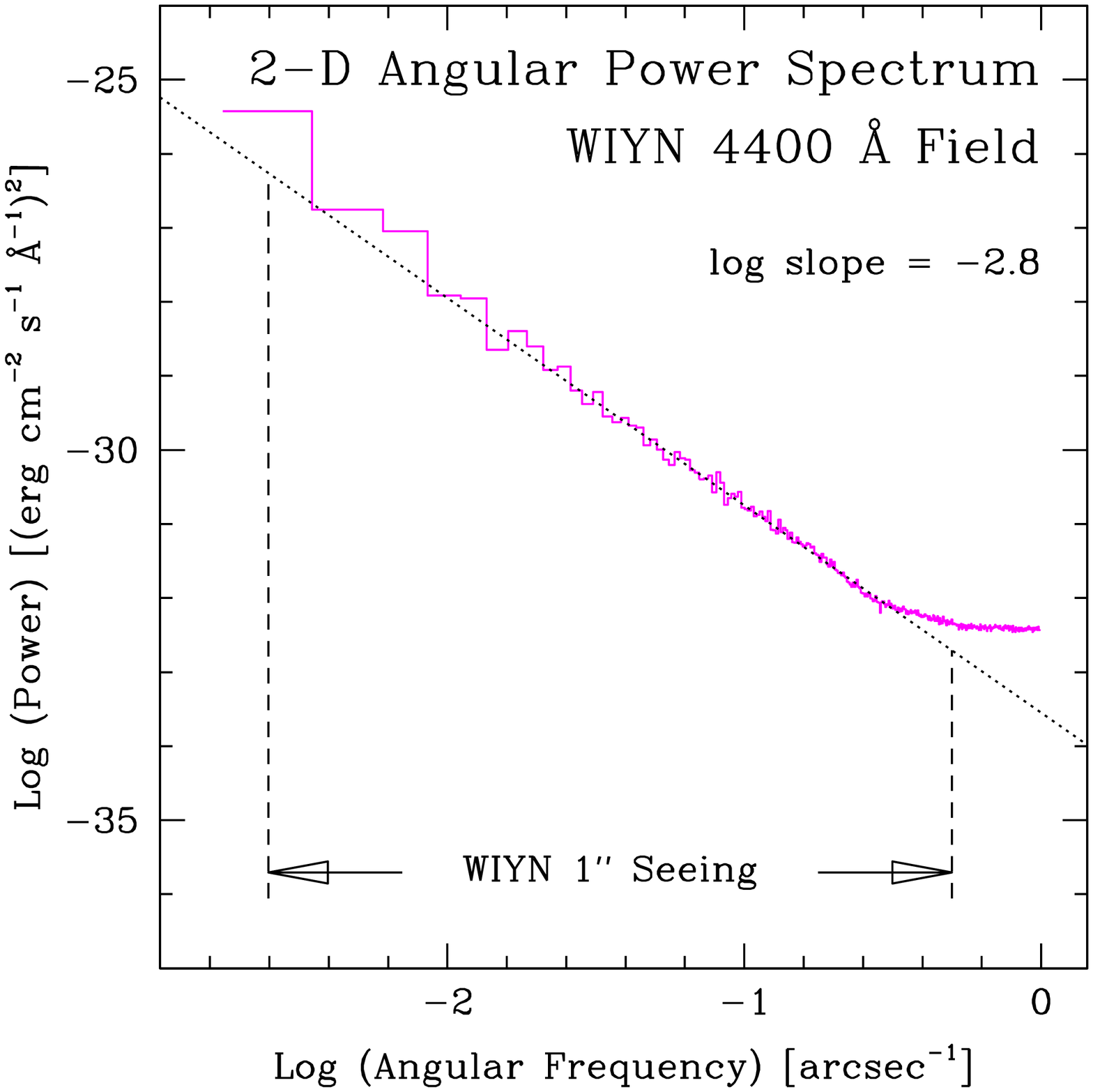}
\plottwo{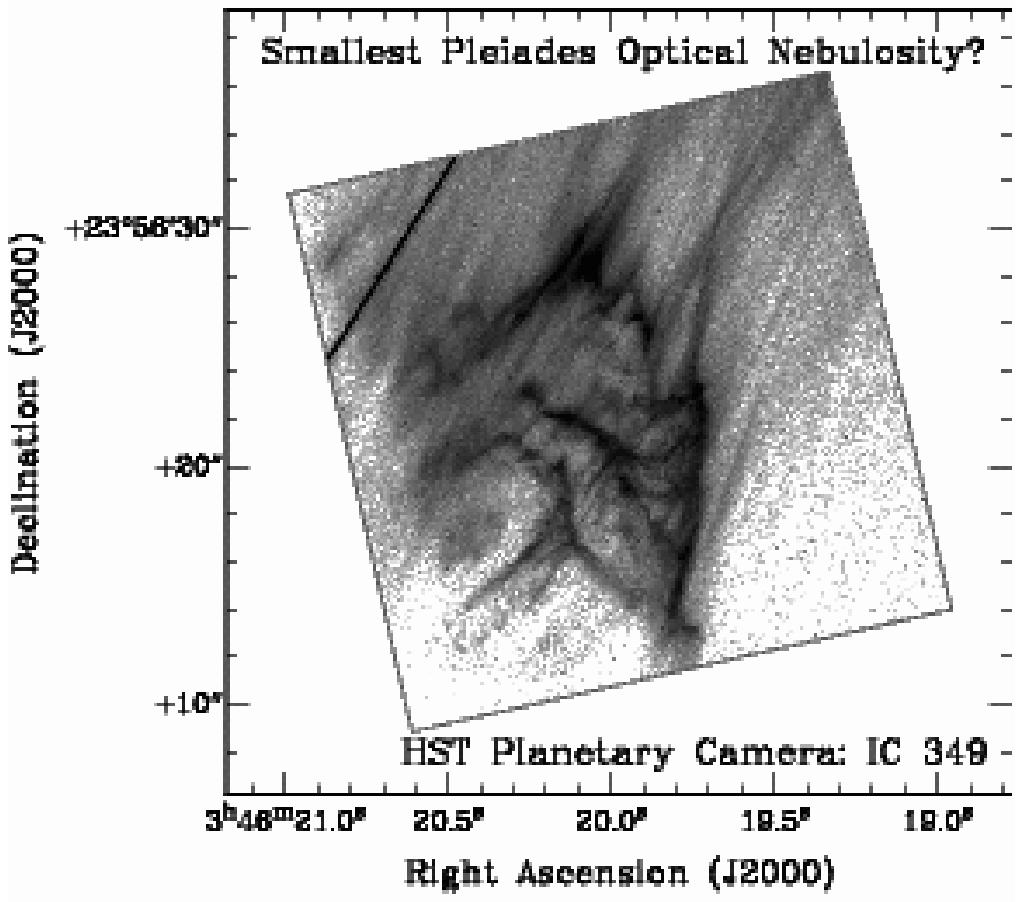}{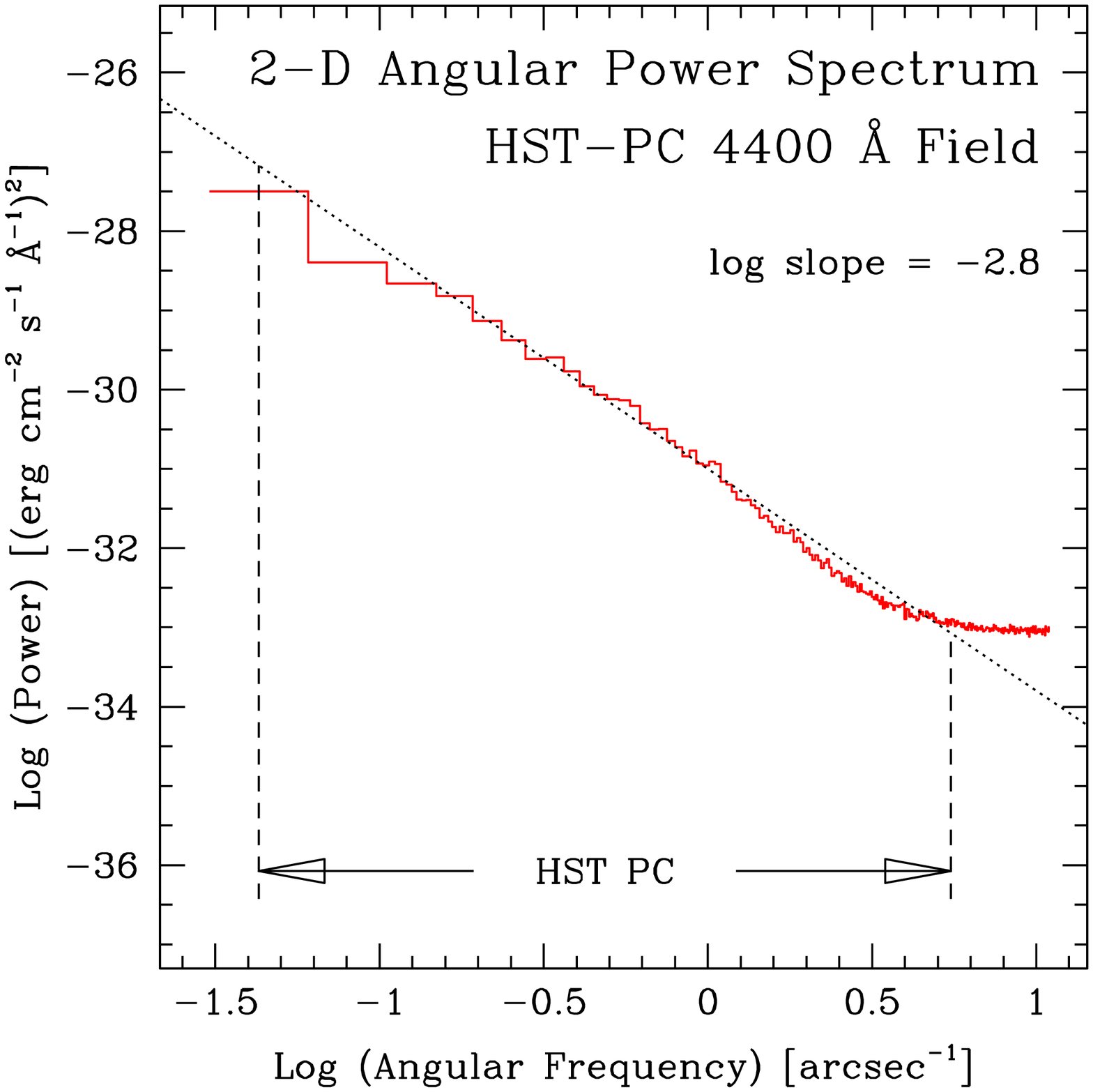}
\caption
{
Pleiades optical nebula, negative logarithmic intensity scale.  
{\itshape Top left:\/} Wisconsin-Indiana-Yale-NOAO (WIYN) 3.5m image of nebular
fine structure east of the 4th-magnitude star Merope, extended to indicate the
position of the star and of the panel below.  
Image courtesy of C. J. Conselice and J. S. Gallagher.  
{\itshape Bottom left:\/} {\sl Hubble Space Telescope} Planetary Camera 
(HST-PC) image of Barnard's Merope Nebula, IC 349 \citep{hs01}, $30''$ south of
Merope.  Image courtesy of T. Simon.
{\itshape Right panels:\/} 2-D angular power spectra derived from the images
at left.  
}
\label{Fig:wiyn}
\end{figure}

\begin{figure}[!ht]
\plottwo{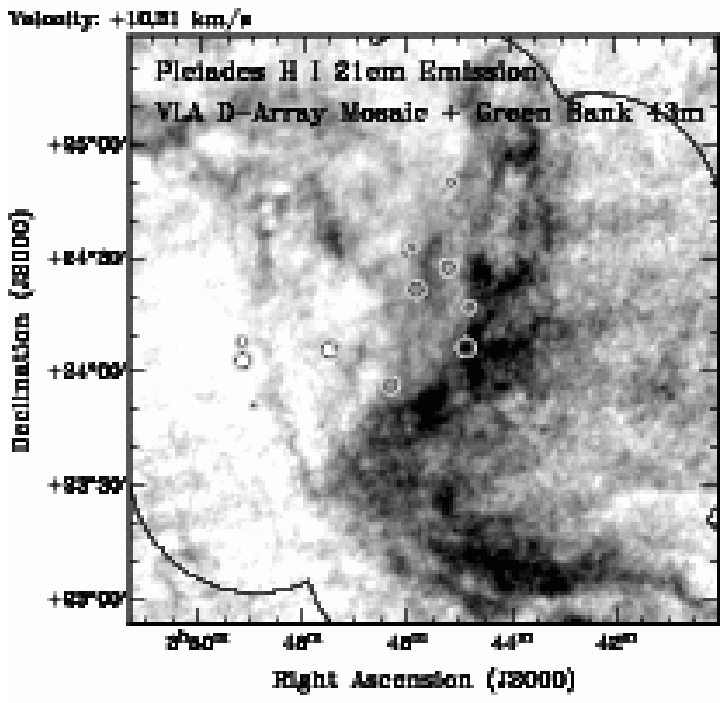}{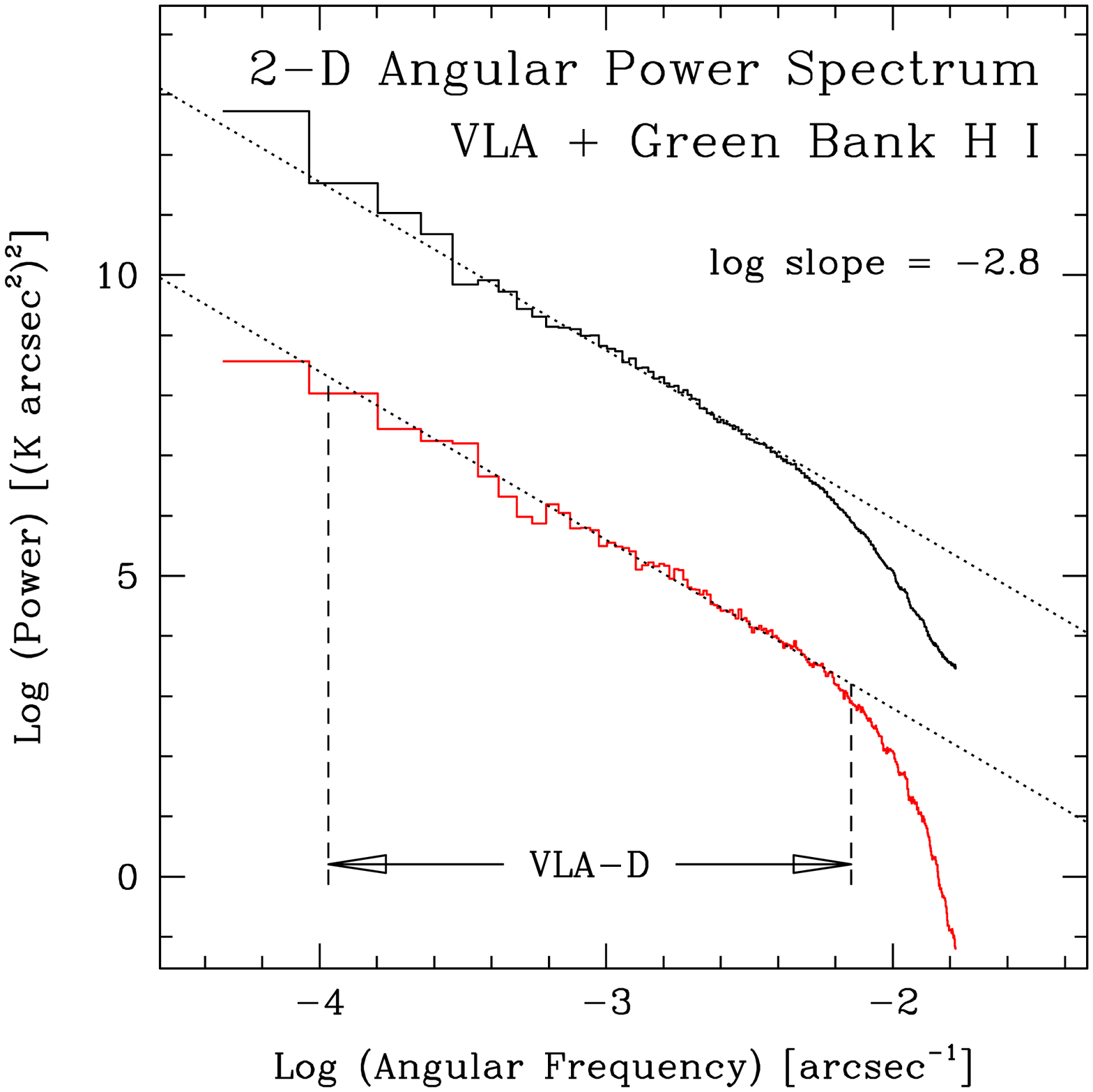}
\plottwo{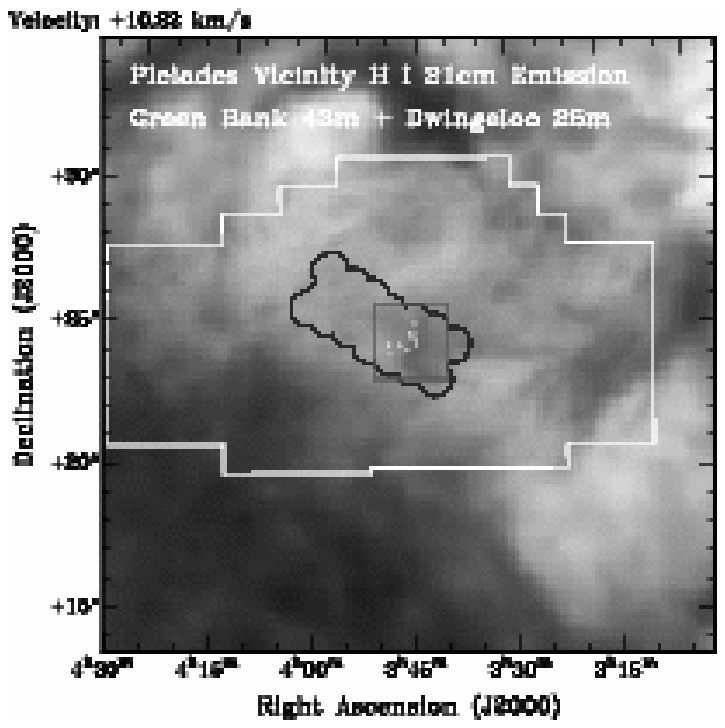}{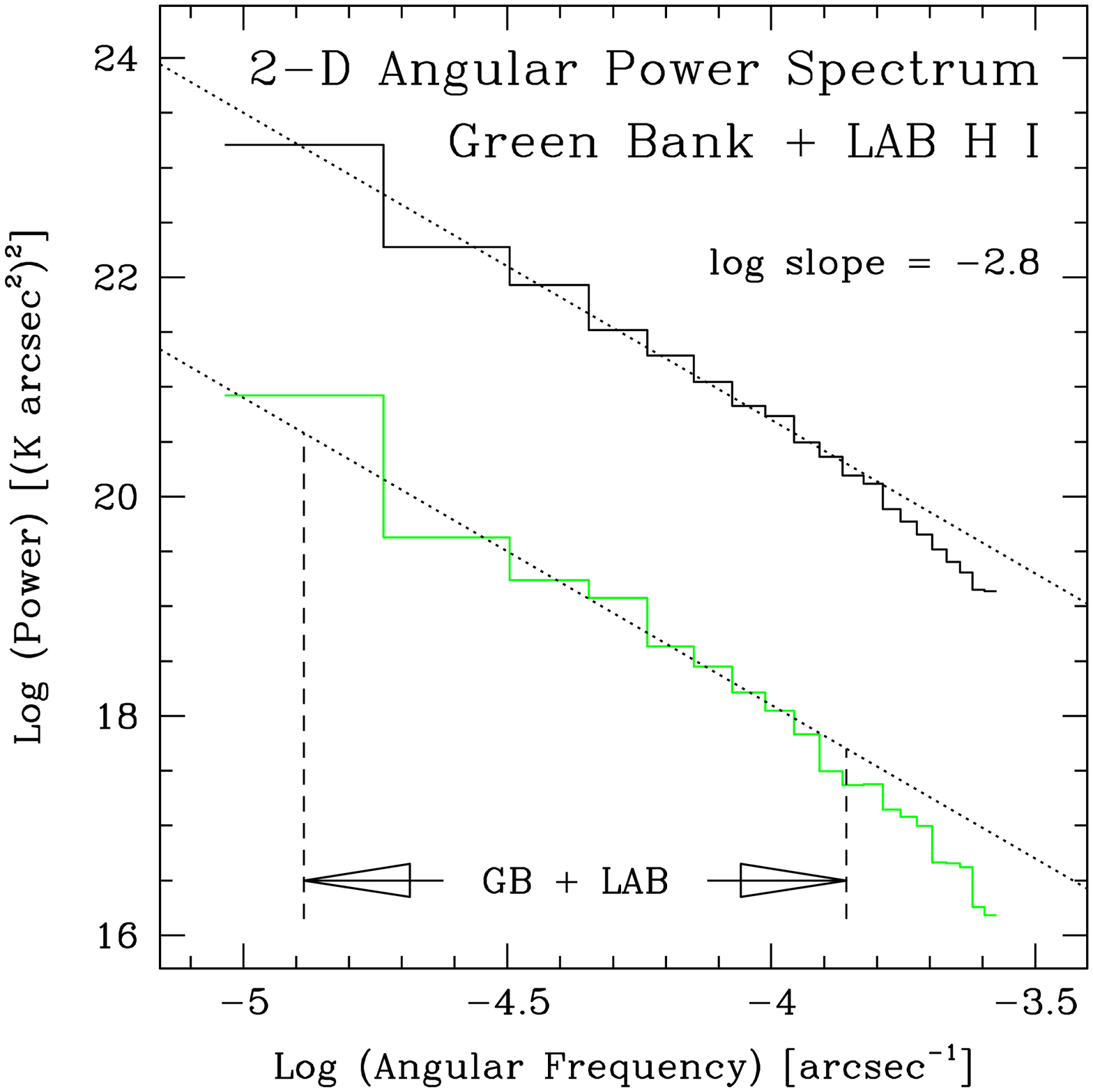}
\caption
{
Pleiades \hi\ 21cm-line emission, negative scale.  
Bright stars and Schmidt mosaic boundaries are shown for reference.
{\itshape Top left:\/} 192-field mosaic using the D-configuration Very Large 
Array (VLA-D), with short spacings provided by the Green Bank 43m telescope
\citep{ghn95}.  Linear intensity scale.  
{\itshape Bottom left:\/} Full Green Bank map (white contour) padded to 128x128
GB pixels with Dwingeloo/LAB 25m data \citep{lab05} for FFT analysis.  VLA 
area is small square.  Logarithmic intensity scale.
{\itshape Right panels:\/} Angular power spectra.  
The lower spectra are for single channels with $v_{lsr} = 
+10~{\rm km\,s^{-1}}$; 
the upper spectra are integrated over the \hi\ line.
}
\label{Fig:hi}
\end{figure}

Here I present optical and \hi\ maps of the ISM associated with the Pleiades,
where dust grains in a passing cloud are illuminated as a reflection nebula
\citep{ga84,w84,b87}.  Since the cluster is nearby ($\sim 130$~pc;
\citealt{psg05}) and there is minimal dust in front of the nebula \citep{c87},
optical features are easily related to specific spatial scales.  The nebula's
rich optical structure is illustrated in {\bf Figures \ref{Fig:schmidt} \&
\ref{Fig:wiyn}}.  \hi\ maps must be interpreted with more caution, since some
emission may lie in the background, but the emission shown in {\bf Figure
\ref{Fig:hi}} has the same velocity as interstellar absorption associated with
the nebula and matches the dust distribution.

\begin{figure}[!ht]
\plotone{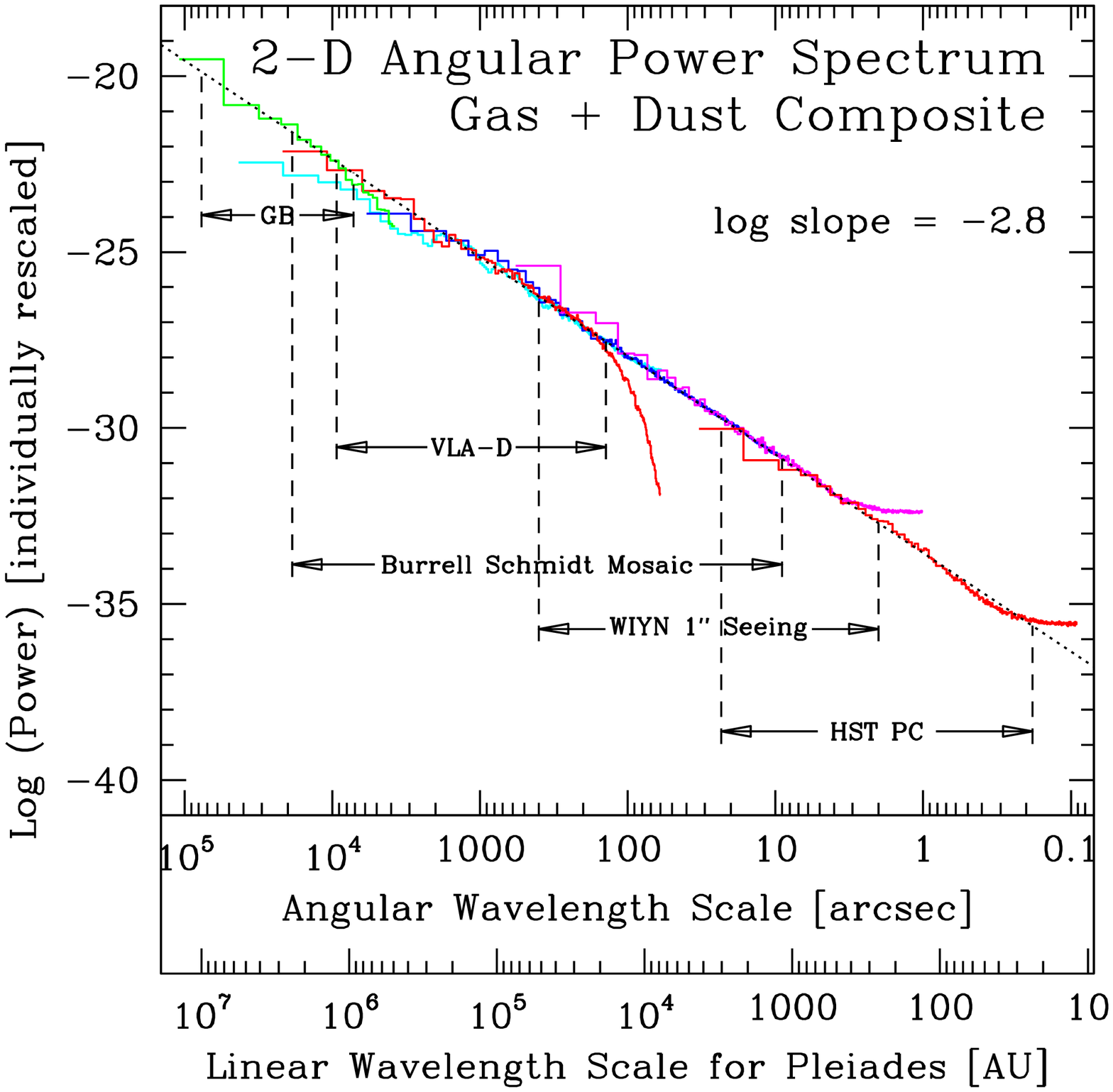}
\caption
{
Composite of the optical and \hi\ angular power spectra shown in Figures 
\ref{Fig:schmidt}-\ref{Fig:hi}.  The relative scalings have been adjusted
to fit a uniform power law, but the slope of each spectrum is unaltered.
The dashed lines mark ranges of measurement for each data set.
``Wavelength'' = 1/spatial frequency.  The
maximum wavelength of any image is the image width, and the minimum is either
twice the beam width or twice the pixel width, whichever is greater.
}
\label{Fig:big_spectrum}
\end{figure}

\section{Analysis}

Each image was run through a 2-D Fast Fourier Transform (FFT) algorithm
\citep{nr88}, and a map of the FFT modulus was computed as the quadrature sum
of the real and imaginary transform components.
The modulus values were then binned by ``radial frequency'' $f_r \equiv
\sqrt{{f_x}^2+{f_y}^2}$, and the 2-D angular power spectrum $P(f_r)$ was
constructed as the square of the median modulus in each $f_r$ bin.  The median
statistic is robust against most artifacts that arise from ``wraparound
discontinuities'' between the right and left or top and bottom image edges.
However, the 2 or 3 lowest-frequency bins may still have excess power in some
cases, particularly for the WIYN data.
Image apodization alleviated this problem for the \hi\ maps but did not help
with the WIYN data; the Schmidt and HST images produced minimal edge artifacts.
Stars were also removed from the WIYN image prior to the FFT, so its power
spectrum is quite clean aside from edge effects.
No significant stars were present in the HST image.  Stars were not removed
from the Schmidt mosaic, because the total map flux is dominated by the
nebulosity near the brightest stars; however all saturated pixels were replaced
by neighbor interpolation.

The power spectra are shown individually in {\bf Figures \ref{Fig:schmidt} -
\ref{Fig:hi}} and together as a composite spectrum in {\bf Figure
\ref{Fig:big_spectrum}}.  Angular frequency is measured in cycles per
arcsecond.  The dashed lines indicate valid ranges of measurement, particularly
where the resolution limits of different data sets occur; near the resolution
frequency, photon noise flattens the WIYN and HST spectra, and \hi\ power drops
precipitously.  Over 62\% of the large-scale \hi\ map is Dwingeloo data, so the
$1^\circ$ LAB Nyquist limit sets the resolution frequency for that spectrum.

\section{Discussion}

All power spectra in {\bf Figures \ref{Fig:schmidt} - \ref{Fig:big_spectrum}}
are plotted against a power law with the same slope of $-2.8$, which was
adopted from visual inspection.
Within the dashed lines, most spectra are remarkably consistent with this power
law.  This includes the Schmidt spectra at scales $\la 1000''$ ($\log{f_r} \ga
-3$), despite the presence of stars in the image.  The consistency between the
WIYN and Schmidt spectra here supports both being dominated by nebulosity.  At
scales $\ga 1000''$, the Schmidt spectrum deviates from the power law, probably
because the stellar illumination pattern is becoming more important than dust
density structure.  For scales larger than the cluster core, the spectrum
flattens in the absence of additional illumination structure.  Similar
``illumination bias'' behavior is seen in power spectra of the smooth nebular
models of \citet{gn03b}, at both optical and far-infrared wavelengths.  The FIR
model results indicate that, unfortunately, 100~$\mu$m maps from the {\sl
Infrared Astronomical Satellite (IRAS)} cannot be used to study multiscale
structure in the Pleiades, because the $5\arcmin$ {\sl IRAS} beam is too close
to the illumination scale.

The HST results are surprising.  While some of the long, parallel streamers in
the map are probably part of the larger nebular structure visible in the WIYN
image, the bright structure in the HST map is from IC~349, an intense 20$''$
clump whose dynamical relation to the larger Pleiades nebulosity is debated
\citep{be99,hs01,w03}.  IC~349 has the appearance of a lumpy, limb-brightened,
optically-thick cloud, while most Pleiades optical nebulosity resembles smooth,
optically-thin filaments.  Thus it seems unlikely that IC~349 is typical of
nebular structure elsewhere in the Pleiades at these scales.  Yet the HST power
spectrum is quite similar to the WIYN and Schmidt spectra.  It is hard to
understand this result unless the power spectrum is relatively insensitive to
morphology.

The \hi\ power spectra are preliminary, as they have not yet been corrected for
possible noise contamination.  But if noise is not significant, as is probably
true for at least the single-dish data, then the \hi\ and dust slopes agree
well enough to suggest a single power law for both, even at scales larger than
that of the Pleiades nebulosity.  This could indicate that the Pleiades ISM
structure is the same as the more general ISM, which is consistent with the
nebula merely being a chance illumination of ordinary interstellar material.
On the other hand, there is no apparent difference in the slope of
single-channel \hi\ emission vs.\ that integrated over all relevant velocities,
which does not agree with the theoretical predictions of \citet{lp00} for the
general ISM.  In any case, when combined, all of the Pleiades power spectra
appear consistent with a single power law over 5 orders of magnitude in angular
and physical scale.

\section{The Next Step: \hi\ Self-Absorption Structure}

Near the Galactic plane, where dust and \hi\ emission are more confused, \hi\
self-absorption (HISA) may serve as a useful cold gas tracer (e.g.,
\citealt{g05}).  A HISA power spectrum investigation is underway.

\acknowledgements 

I would like to thank my long-term Pleiades collaborators Ken Nordsieck and
Mark Holdaway.  Work on this project was supported by the Natural Sciences and
Engineering Research Council of Canada and the U.S. National Science
Foundation.


\end{document}